\documentstyle[11pt,aas2pp4,psfig]{article}
\begin{document}

\lefthead{Psaltis, Belloni, \& van der Klis}
\righthead{Quasi-Periodic Oscillations}

\title{Correlations in Quasi-Periodic Oscillation and
Noise Frequencies Among Neutron-Star and Black-Hole
X-ray Binaries}

\author{Dimitrios Psaltis}

\affil{Harvard-Smithsonian Center for
Astrophysics, 60 Garden St., Cambridge, MA 02138,
U.S.A.; dpsaltis@cfa.harvard.edu}

\author{Tomaso Belloni, and Michiel van der Klis}

\affil{Astronomical Institute ``Anton Pannekoek'',
University of Amsterdam and Center for High Energy
Astrophysics, Kruislaan 403, NL-1098 SJ Amsterdam,
The Netherlands;\\ tmb, michiel@astro.uva.nl}

 \begin{abstract}
 We study systematically the $\simeq 0.1-1200$~Hz
quasi-periodic oscillations (QPOs) and broad noise
components observed in the power spectra of non-pulsing
neutron-star and black-hole low-mass X-ray binaries.
We show that among these components we can identify
two, occurring over a wide range of source types and
luminosities, whose frequencies follow a tight
correlation. The variability components involved in
this correlation include neutron-star kilohertz QPOs
and horizontal-branch oscillations, as well as
black-hole QPOs and noise components. Our results
suggest that the same types of variability may occur
in both neutron-star and black-hole systems over
three orders of magnitude in frequency and with
coherences that vary widely but systematically. 
Confirmation of this hypothesis will strongly
constrain theoretical models of these phenomena and
provide additional clues to understanding their
nature. 
 \end{abstract}


\vspace{0.5cm}

\begin{center}
To appear in the {\em Astrophysical Journal}
\end{center}

\newpage

\section{INTRODUCTION}

Fast quasi-periodic oscillations (QPOs) in the
X-ray brightness of neutron-star and black-hole
X-ray binaries provide a useful probe into the
inner accretion flows around such compact objects.
Since the original discovery of $\simeq 20-50$~Hz
QPOs in the luminous neutron-star binary GX~5$-$1
(van der Klis et al.\markcite{Ketal85} 1985), a
variety of additional QPOs as well as broad noise
components have been observed with properties that
depend on the spectral state of the sources (see
van der Klis\markcite{vdK95}\markcite{vdK98} 1995,
1998 for reviews).

The Z sources, which are luminous neutron-star
low-mass X-ray binaries (Hasinger \& van der
Klis\markcite{HK89} 1989), typically show four
distinct types of QPOs.  In current nomenclature,
these are the $\simeq 5-20$~Hz normal branch
oscillation (NBO), the $\simeq 15-60$~Hz horizontal
branch oscillation (HBO; see van der Klis 1989 for a
review of these low frequency QPOs), and the $\simeq
200-1200$~Hz kilohertz QPOs that typically occur in
pairs (van der Klis et al.\ 1996). The atoll sources,
which are less luminous neutron-star low-mass X-ray
binaries, typically show $\sim 500-1250$~Hz kilohertz
QPOs that occur in pairs (Strohmayer et al.\ 1996), as
well as $\sim 20-60$~Hz QPOs and broad noise
components that have been identified as possibly
similar to horizontal branch oscillations (see, e.g.,
Hasinger \& van der Klis 1989; Homan et
al.\markcite{Hetal98} 1998).  All of these QPOs have
centroid frequencies that increase with inferred mass
accretion rate. In several atoll sources, nearly
coherent $\simeq 300-600$~Hz oscillations have also
been detected during thermonuclear Type~I X-ray bursts
(see, e.g., Strohmayer et al.\markcite{Setal96} 1996). 

The phenomenology of QPOs in black-hole X-ray binaries
has not been developed to the same extent so far. 
Several low-frequency ($\sim 10^{-1}-10$~Hz) QPOs have
been detected with frequencies that depend on inferred
mass accretion rate (see, e.g., Chen, Swank, \&
Taam\markcite{CST97} 1997; Morgan, Remillard, \&
Greiner\markcite{MRG97} 1997), as well as three cases
of QPOs with higher frequencies, two of which may or
may not be constant ($\simeq 67$~Hz in GRS~1915$+$105: 
Morgan, Remillard, \& Greiner\markcite{MRG97} 1997; 
$\simeq 300$~Hz in GRO~1655$-$40:  Remillard et
al.\markcite{Retal99a} 1999a; $\sim 160-220$~Hz in
XTE~J1550-564:  Remilard et al.\markcite{Retal99b}
1999b). Broad noise components are also prominent in
the power-density spectra of black-hole X-ray binaries
and show many similarities with those of neutron-star
sources (van der Klis 1994a, 1994b; Wijnands \& van
der Klis\markcite{WV99} 1999). 

A large variety of theoretical models have been
proposed for the different QPOs in neutron-star and
black-hole systems.  The rms amplitudes of almost all
QPOs increase with increasing photon energy up to
$\simeq 10-30$~keV and their frequencies usually
depend strongly on mass accretion rate (see, however,
Jonker, van der Klis, \& Wijnands 1999).  For these
and other reasons, such QPOs are thought to originate
close to the compact objects and their frequencies
have been identified with various characteristic
frequencies in the inner accretion flows.  For
example, theoretical models attribute some of the
observed QPOs to Keplerian orbital frequencies in the
disk (e.g., Alpar et al.\markcite{Aetal92} 1992;
Miller, Lamb, \& Psaltis\markcite{MLP98} 1998), to the
beat of such frequencies with the stellar spin (e.g.,
Alpar \& Shaham\markcite{AS85} 1985; Lamb et
al.\markcite{Letal85} 1985;  Miller et al.\ 1998), to
radiation-hydrodynamic (Fortner, Lamb, \&
Miller\markcite{FLM89} 1989; Klein et
al.\markcite{Ketal96} 1996) or oscillatory disk and
stellar modes (e.g., Nowak \& Wagoner\markcite{NW91}
1991; Bildsten \& Cutler\markcite{BC95} 1995;
Kanetake, Takauti, \& Fukue\markcite{KTF95} 1995;
Alpar \& Yilmaz\markcite{AY98} 1998; Titarchuk,
Lapidus, \& Muslimov\markcite{TLM98} 1998), to general
relativistic effects (Ipser\markcite{I96} 1996; Stella
\& Vietri\markcite{SV98} 1998, 1999), etc.  Some of
these models depend explicitly on the existence of a
hard surface or a large-scale magnetic field and
therefore are valid only for QPOs in neutron-star
systems, whereas others are applicable only to
black-hole systems or to both.

The detection of several QPOs at the same time in a
single source provides multiple probes to the inner
accretion flow around the compact object.  Such a
simultaneous detection is often attributed to actions
of distinct mechanisms affecting the X-ray brightness
of the source simultaneously in different ways,
whereas in other models all QPOs correspond to
different modes of the same fundamental mechanism
(e.g., Titarchuk \& Muslimov\markcite{TM97} 1997).
When several types of QPO are observed in the X-ray
brightness of a system simultaneously, their
frequencies, which commonly increase with mass
accretion rate, are often tightly correlated (see,
e.g., van der Klis et al.\ 1996; Ford \& van der
Klis\markcite{FK98} 1998; Psaltis et al.\ 1998, 1999).
In some cases, these correlations appear to depend
only weakly on the other properties of the sources.
For example, the frequencies of the upper and lower
kHz QPOs, as well as the frequencies of the upper kHz
QPO and the HBO are consistent with following very
similar relations in all Z sources (Psaltis et al.\
1998, 1999).

Here we study the systematics of QPOs and peaked broad
noise components observed in non-pulsing neutron-star
and black-hole low-mass X-ray binaries. We find tight
correlations among them. In particular, we find an
indication that two types of variability, showing up
in some systems as an HBO and a kHz QPO, may occur, at
widely different coherences and over a wide range of
frequencies, in both neutron-star and black-hole
systems. We caution, however, that the striking
correlations we report here might be an artifact
produced by the accidental line-up of multiple,
independent correlations. If future observation
confirm our results, this will strongly constrain
theoretical models of QPOs in neutron-star and
black-hole systems.

\section{IDENTIFICATION OF QPOs IN\\ NEUTRON-STAR AND 
BLACK-HOLE SYSTEMS}

Quasi-periodic oscillations and peaked broad noise
components have been detected in both neutron star and
black hole systems. In this section we discuss
observations of such systems in which at least one QPO
has been detected together with one more peaked
variability component, and attempt to identify the
various power-spectral components with QPOs and noise
components in other sources on the basis of,
primarily, their frequencies.  We restrict our study
to phenomena with frequencies $\gtrsim 0.1$~Hz to
avoid the very complicated timing behavior of the
microquasars at these low frequencies (see, e.g.,
Remillard et al.\ 1999a). We do not consider in our
study the peculiar $\simeq 1$~Hz QPO in the dipper
4U~1323$-$62 (Jonker et al.\ 1999) that has properties
very different compared to all other QPOs in
non-pulsing neutron-star sources. We will also exclude
from our study QPOs observed in accretion powered
pulsars, most of which are thought to be strongly
magnetic ($\sim 10^{12}$~G) and hence to have
different inner accretion flow properties than the
non-pulsing neutron star sources. The QPO and noise
properties in the only known millisecond accretion
powered pulsar, SAX~J1808.4-3658, have been compared
to those of the non-pulsing sources by Wijnands \& van
der Klis (1999)  and are also excluded in the present
study.

\begin{figure*}[ht]
 \centerline{
\psfig{file=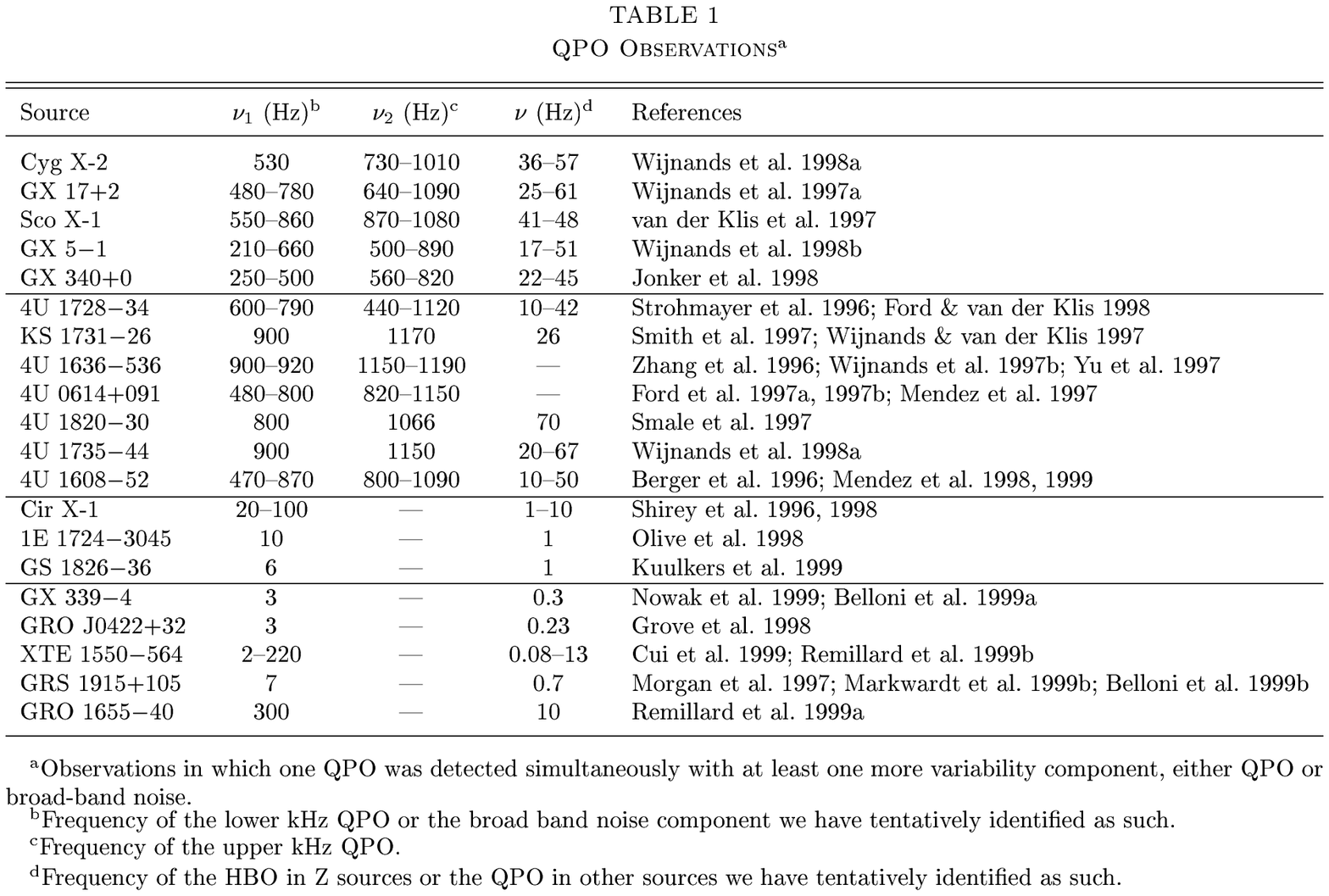,angle=0,height=270pt,width=400pt}}
\end{figure*}

\subsection{Neutron-Star Sources}

{\em Z sources.---\/}In the power spectra of Z
sources, the identification of QPOs and noise
components is unambiguous. For example, four
distinct QPO peaks that are not harmonically
related are often observed simultaneously in
Sco~X-1 (van der Klis et al.\ 1996). In this as
well as the other Z sources, these QPO peaks are
identified as NBOs, HBOs, or kHz QPOs, based on
their frequencies and occurrence in different
spectral branches. Figure~1 shows a
power-density spectrum of Sco~X-1 (based on the
data described in van der Klis et al.\ 1997), in
which an HBO and its harmonic, as well as a
lower and an upper kHz QPO are evident (notice
that frequency shifts were applied to the power
spectra shown in Fig.\,1).  In the five Z
sources in which HBOs have been observed
simultaneously with two kHz QPOs, the frequency
of the HBO is tightly correlated to the
frequency of the upper kHz QPO (Wijnands et al.\
1998a, 1998b; Homan et al.\ 1998;  see also
Stella \& Vietri 1997; Psaltis et al.\ 1999). 
At the same time, the frequencies of the lower
and upper kHz QPOs are also tightly correlated,
in a way that is well described by a simple empirical
power-law relation (Psaltis et al.\ 1998). Both
correlations are consistent with being very similar
in all Z sources. In Sco X-1, which is the only
Z source that simultaneously shows NBOs and kHz
QPOs, the frequencies of these QPOs are also
tightly correlated (van der Klis et al.\ 1997). 
Here, in discussing the QPOs detected in Z
sources, we shall use {\em RXTE} data that have
been previously published (see Table~1).

The upper kHz QPO, HBO, and NBO frequencies detected
in the above Z sources are plotted in Figure~2 against
the lower kHz QPO frequency (red points). The HBO
frequency $\nu_{\rm HBO}$ appears to be well
correlated to the frequency of the lower kHz QPO
$\nu_1$, as expected given their previously known
common dependence on the frequency of the upper kHz
QPO. When $\nu_1 \lesssim 550$~Hz, the data points for
the Z sources are consistent with the empirical
relation $\nu_{\rm HBO}\simeq (42\pm
3$~Hz$)(\nu_1/500$~Hz$)^{0.95\pm0.16}$, in which the
normalization constant depends on the peak separation
of the kHz QPOs, which is very similar between sources
(compare Psaltis et al.\ 1998, 1999; when $\nu_1
\gtrsim 550$~Hz, $\nu_{\rm HBO}$ increases slower with
$\nu_1$).  Figure~2 shows this relation (dashed line),
extrapolated by more than two orders of magnitude
towards lower frequencies.

\begin{figure}[ht]
\centerline{
\psfig{file=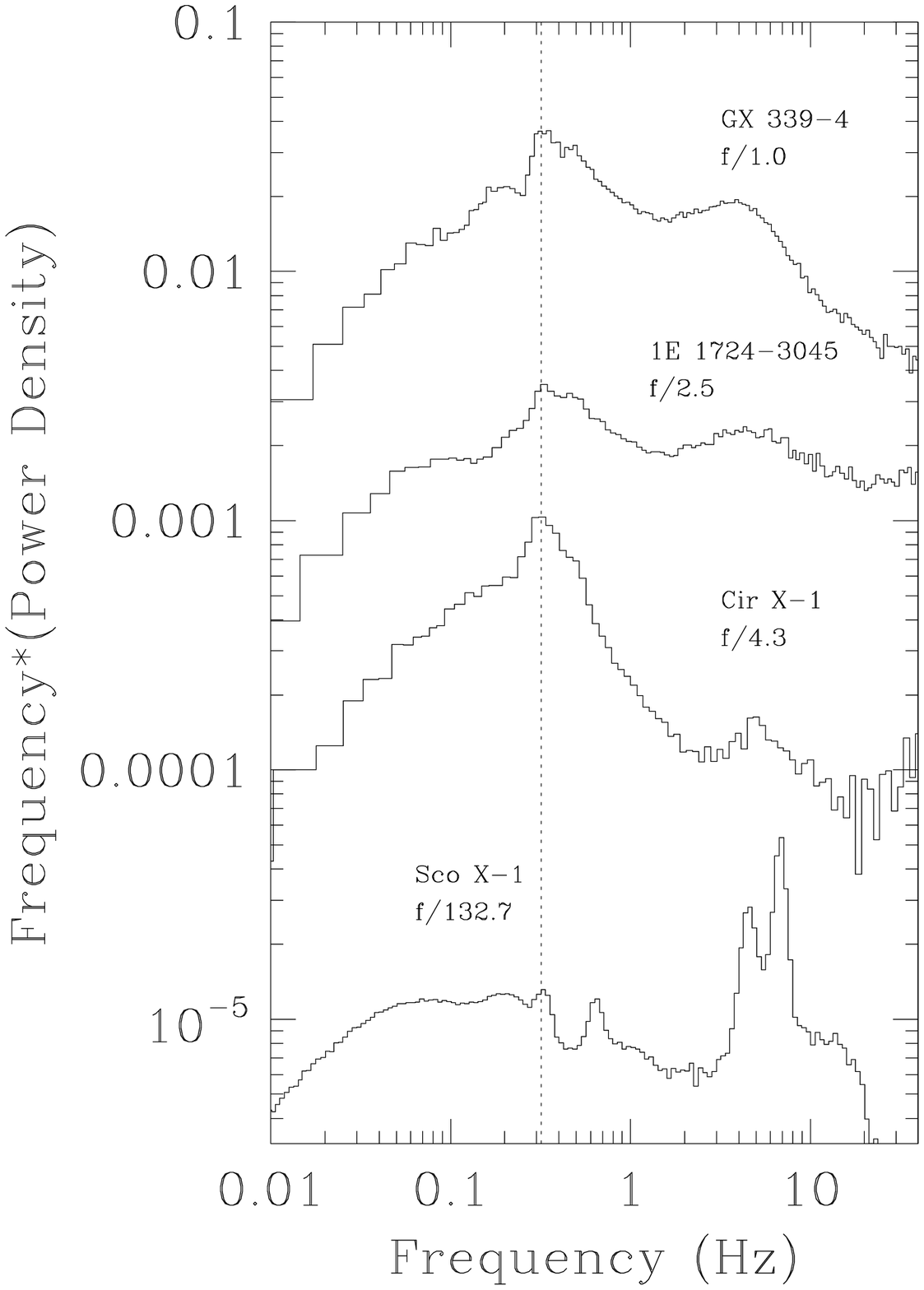,angle=0,height=300pt,width=230pt}}
\end{figure}

{\em Atoll sources.---\/}In most atoll sources in
which QPOs have been detected, kHz QPOs can be easily
identified when at high frequencies, because of these
high frequencies and their common occurrence in pairs.
However, when their frequencies become low, the
coherence of the QPO decreases and confusion becomes
more likely, especially with other $\sim 100$~Hz
components of the power spectra (Ford \& van der Klis
1998). As in the case of the Z sources, the
frequencies of simultaneously detected lower and upper
kHz QPOs are tightly correlated. They are consistent
with following empirical power-law relations similar
to the one followed by the Z sources (Psaltis et al.\
1998). Here, we shall use kHz QPO data for atoll

 \figcaption[]
 {Examples of power spectra of low-mass X-ray
binaries, in which more than one QPO or broad-band
noise components are detected. The individual power
spectra were shifted along the vertical axis for
clarity and along the horizontal axis, by the amounts
displayed, for the low-frequency QPOs to be aligned
(dotted line). The sample of sources includes a
black-hole candidate (GX~339$-$4; Belloni et al.\
1999a), an X-ray burster (1E~1724$-$3045; Olive et
al.\ 1998), a luminous neutron-star (Cir~X-1;  Shirey
1998), and a Z source (Sco~X-1). The continuum in the
power spectrum of Sco~X-1 at high frequencies is
affected by instrumental effects.}

\vspace*{0.3cm}

\noindent sources that have been obtained by {\em
RXTE\/} and have been previously published (see
Table~1). Figure~2 shows the upper kHz QPO frequencies
detected in atoll sources plotted against the lower
kHz QPO frequencies (blue points near the top of the
graph).

Single QPOs at frequencies $\simeq 20-70$~Hz have
occasionally been detected in some atoll sources
simultaneously with the kHz QPOs (here we use the {\em
RXTE\/} data on 4U~1820$-$30:  Smale, Zhang, \&
White\markcite{SZW97} 1997;  KS~1731$-$26:  Wijnands,
\& van der Klis\markcite{WK97} 1997; 4U~1608$-$52:  
M\'endez et al.\markcite{Metal99} 1999; 4U~1728$-$34:
Ford \& van der Klis 1998; and 4U~1702$-$429:
Markwardt, Strohmayer, \& Swank\markcite{MSS99a}
1999a; see also Table~1). Given their low frequencies,
it has been suggested that these QPOs can be
identified with the HBO.  In 4U~1728 $-$34, these QPOs
have frequencies that increase with those of the kHz
QPOs; during some of the observations, the correlation
between the $\simeq 20-70$~Hz QPO and upper kHz QPO
frequencies was similar to the one followed by the Z
sources, whereas during one other observation the HBO
frequencies were systematically lower by a factor of
$\simeq 2$ (Ford \& van der Klis 1998). In
4U~1735$-$44, {\em two\/} low-frequency QPOs have been
simultaneously detected with the kHz QPOs (Wijnands et
al.\markcite{Wetal98a} 1998a), reminiscent of the
simultaneous detection of HBOs and NBOs in the Z
sources. In 4U~1702$-$429, multiple $\sim 10-100$~Hz
QPOs are occasionally detected, but not simultaneously
with kHz QPOs (Markwardt et al.\ 1999). Finally, in
4U~1820$-$30, a $\simeq 7$~Hz QPO is occasionally
detected, but not simultaneously with kHz QPOs
(Wijnands, van der Klis, \& Rijkhorst\markcite{WKR99}
1999).

Figure~2 shows the frequencies of the $\simeq
20-70$~Hz QPOs observed in atoll sources plotted
against the frequencies of the simultaneously-detected
lower kHz QPOs (blue points in the middle of the
graph\footnote{Note that in plotting most of the data
points of 4U~1728$-$34 we have inferred the frequency
of the lower kHz QPO, which was not always detected,
by subtracting the 363~Hz frequency of the
oscillations observed during type~I X-ray bursts from
this source from the frequency of the detected upper
kHz QPO.}). The QPO in 4U~1820$-$30, the higher of the
two low-frequency QPOs in 4U~1735$-$44, and most of
the data points of 4U~1728$-$34 follow a trend very
similar to the correlation between the HBO and lower
kHz QPO frequencies observed in Z sources. On the
other hand, some of the data points of 4U~1728$-$34 as
well as the data points of 4U~1608$-$52 and
4U~1702$-$429 follow a similar trend but at
systematically lower frequencies (see also Ford \& van
der Klis 1998). Finally, the QPO in KS~1731 $-$26 and
the lower of the two low-frequency QPOs in
4U~1735$-$44 do not follow either of the above two
correlations, showing a hint of similarity with the
NBO observed in Sco~X-1.

{\em Other neutron-star sources.---\/}Cir~X-1 is a
probable neutron-star low-mass X-ray binary (Tennant,
Fabian, \& Shafer\markcite{TFS86} 1986) that was
initially labeled as a black-hole candidate
(Toor\markcite{T77} 1977; Samimi et
al.\markcite{Setal79} 1979) and shows similarities to
black holes as well as to both Z and atoll sources
(Oosterbroek et al.\markcite{Oetal95} 1995;  
Shirey\markcite{S98} 1998). At intermediate
luminosities, one QPO is detected in this source at
frequencies varying between $\simeq 1-10$~Hz, together
with a broader power-spectral component with a
centroid frequency varying between $\simeq 10-100$~Hz
(Shirey et al.\markcite{Setal97}\markcite{Setal98}
1996, 1998; Shirey 1998; see Figure~1; see also
Tennant [1987] for a possible detection of a $\simeq
200$~Hz broad component simultaneously with a $\simeq
12$~Hz QPO using {\rm EXOSAT\/}).

The dependence on countrate of the frequencies of the
narrow $\simeq 1-10$~Hz QPO and of the broad $\simeq
10-100$~Hz component as well as their frequency ratio
are similar to those of the HBO and lower kHz QPO seen
in Z sources (Shirey et al.\ 1996, 1998, 1999; Shirey
1998). Therefore, in spite of the broad component
being much less coherent than the lower kHz QPOs seen
in other sources, we tentatively identify these two
components seen in Cir~X-1 with the HBO and the lower
kHz QPO seen in Z sources (see also Shirey et al.\
1999). Figure~2 shows the frequency of the narrow QPO
in Cir~X-1, which we have identified with the HBO,
plotted against the frequency of the broad noise
components, which we have identified with the lower
kHz QPO (cyan points). The data points of Cir~X-1 are
consistent ($\chi^2$/d.o.f. $\simeq 1$) with the
relation between the HBO and lower kHz QPO frequencies
observed in Z sources and extrapolated toward lower
frequencies, strengthening our identification. 

A similar combination of a $\simeq 1$~Hz QPO and a
$\simeq 5-10$~Hz broad power-spectral component has
also been observed in the X-ray bursters GS~1826$-$36
(Kuulkers et al.\markcite{Ketal99} 1999) and
1E~1724 $-$3045 (Olive et al.\markcite{Oetal98} 1998).
Note here that Olive et al.\ (1998) demonstrated that
the power spectrum of 1E~1724$-$3045 at high ($\gtrsim
1$~Hz) frequencies is well described by the sum of two
zero-centered Lorentzians; here we describe the same
power spectra at high frequencies using instead a
power-law continuum and a $\simeq 10$~Hz, broad noise
component (see Fig.\,1) and plot their frequencies in
Figure~2 (cyan points in the lower left part of the
figure).  We again tentatively identify the narrow QPO
and broad noise component observed in these two
bursters with the HBO and lower kHz QPO of the Z
sources as suggested by Figure~2. 

\subsection{Black-Hole Sources}

Low-mass X-ray binaries with black holes show at some
spectral states a narrow $\simeq 0.1-10$~Hz QPO that
is often accompanied with peaks at its harmonics or
even at its first subharmonic. The frequency of this
QPO shows a strong correlation with countrate (see,
e.g., van der Klis 1995 for a review) as well as with
the frequency of the break of the power-density
spectrum that occurs at lower frequencies (Wijnands \&
van der Klis 1999). This correlation is very similar
between black-hole sources and is also similar to the
one between the frequencies of the HBO and the
spectral break in atoll and Z neutron-star sources
(Wijnands \& van der Klis 1999). Based on these
properties, it is therefore tempting to identify the
$\sim 0.1-10$~Hz QPO observed in black-hole sources
with the HBO observed in neutron-star sources. 

No pairs of QPOs with properties similar to those of
the neutron-star kHz QPOs have ever been observed in
any black-hole source. Narrow $\sim 50-300$~Hz QPOs
with frequencies that appear to depend only weakly on
countrate have been observed from some of these
source, two of which are the microquasars
GRS~1915$+$105 (Morgan et al.\ 1997) and GRO~1655$-$40
(Remillard et al.\ 1999a). In other sources, the
low-frequency QPOs are occasionally accompanied by a
$\simeq 1-200$~Hz broad power-spectral component,
similar to those observed in Cir~X-1 (see below). Here
we discuss some timing properties of black-hole
candidates for which QPOs have been reported so far
and attempt to identify their power-spectral
components with similar features seen in neutron-star
sources. Because we are interested in variability
components similar to the neutron-star kHz QPOs, we
will only study the peaked power-spectral components
that occur at frequencies higher than the $\sim
0.1-10$~Hz QPOs. Our study, therefore, complements the
one of Wijnands \& van der Klis (1999), who studied
power-spectral components that occur at frequencies
lower than these QPOs.

{\em GX~339$-$4.---\/}Optical QPOs have been often
observed from this black-hole candidate with
frequencies $\simeq 0.05-0.15$~Hz (Motch et
al.\markcite{Metal83}\markcite{Metal85} 1983, 1985;  
Imamura et al.\markcite{Ietal90} 1990;  
Steiman-Cameron et al.\markcite{SCetal97} 1997).  At
one incidence, when the source was in its low state,
the QPO had a frequency of $\simeq 0.05$~Hz and was
detected simultaneously in X-rays with {\em Ariel 6\/}
(Motch et al.\ 1983). None of these QPOs was detected
simultaneously with another peaked variability
component and hence they cannot be used in the present
study.

{\em SIGMA\/} observations of GX~339$-$4 revealed a
0.8~Hz QPO with a complex, possibly peaked noise
component at frequencies $\simeq 5-10$~Hz, when the
source was in the low state (Grebenev et
al.\markcite{Getal91}\markcite{Getal94} 1991, 1994).  
{\em GINGA\/} observations of the same source at the
low state gave hints of a $\simeq 0.1$~Hz QPO detected
simultaneously with a $\simeq 1-2$~Hz peaked noise
component (Miyamoto et al.\markcite{Metal92} 1992).  
Recent {\em RXTE\/} observations of this source at the
low state, showed a clear $\simeq 0.3$~Hz QPO together
with a $\simeq 3$~Hz peaked noise component (Nowak,
Wilms, \& Dove\markcite{NWD99} 1999; Wijnands \& van
der Klis 1999; Belloni et al.\ 1999a; see also
Fig.\,1). Given the striking similarity, after a shift
of about one decade in frequency, between the power
spectra of GX~339$-$4 observed by {\em RXTE\/} and the
power-spectra of the bursters 1E~1724$-$3045,
GS~1826$-$36, and of Cir~X-1 (see Fig.\,1), as well as
the position of the data points that correspond to
GX~339$-$4 in Figure~2, we tentatively identify these
neutron-star and black-hole power spectral components
as being similar in nature. In Figure~2 we do not
include the data points from the {\em SIGMA\/} and
{\em GINGA\/} observations because the exact
frequencies of the noise components were not reported
for them. However, the power spectra of those two
observations appear also consistent with the general
trend shown in Figure~2.

{\em GINGA\/} observations of GX~339$-$4 in the very
high state showed a $\simeq 5.8-7.4$~Hz QPO
simultaneously with an often peaked noise component
with a photon-energy dependent frequency of $\simeq
1-2$~Hz (Miyamoto et al.\ 1991; Belloni et al.\ 1997;  
Rutledge et al.\ 1999). This noise component is at
frequencies lower than those of the QPO and the
frequencies of the two components follow the
correlation between the frequencies of the break in
the power-spectrum and the QPO found by Wijnands \&
van der Klis (1999) and we therefore also identify
them as such. No other QPO or peaked noise component
was detected at this spectral state.

{\em Cyg~X-1.---\/}Various detections of rather broad
QPOs from Cyg~X-1 have been reported in the
literature. {\em SIGMA\/} observations revealed such a
QPO at frequencies $\simeq 0.05-0.1$~Hz (Vikhlinin et
al.\markcite{Vetal94} 1994), {\em GINGA\/}
observations revealed a $\simeq 1-10$~Hz broad QPO
with spectral properties similar to the narrow QPO
seen in GX~339$-$4 (see Rutledge et al.\ 1999 and
references therein), and {\em RXTE\/} observations
resulted in the detection of a $\sim 3-10$~Hz QPO
during spectral-state transitions (Cui et
al.\markcite{Cetal97} 1997). The correlation between
the frequencies of the QPO and the power-spectral
break in the {\em RXTE\/} observations follow the
correlation between the frequencies of the narrow QPOs
and power-spectral breaks seen in other black-hole and
neutron-star systems (Wijnands \& van der Klis 1999),
thereby suggesting that these QPOs are similar in
nature. The $\simeq 1$~Hz QPO in the {\em GINGA\/}
observations reported by Rutledge et al.\ (1999) is
often accompanied by a $\simeq 10$~Hz broad, peaked
power-spectral component at higher frequencies that
appear to be consistent with the correlations
suggested by Figure~2; the frequencies of the latter
component have not been reported and, therefore,
cannot be used in the present study.

{\em GRO~J0422$+$32.---\/}A prominent QPO at a
frequency $\simeq 0.2-0.3$~Hz was discovered during
the 1992 outburst of this source by {\em SIGMA\/}
(Vikhlinin et al.\markcite{Vetal95} 1995), {\em
BATSE\/} (Kouveliotou, Finger, \&
Fishman\markcite{KFF92} 1992), and {\em OSSE\/} (Grove
et al.\markcite{Getal98} 1998). In the {\em OSSE}
power spectrum a broad noise component at $\simeq
3$~Hz is also evident. Given the striking similarity
between this power spectrum and the power spectra of
neutron-star sources, such as Cir~X-1 and
1E~1724$-$3045 (see, e.g., Olive et al.\ 1998), and
black-hole sources, such as GX~339$-$4, we identify
the QPO and broad noise component in all these sources
as similar in nature. Figure~2 shows that the
frequencies of the QPO and broad noise component in
GRO~J0422$+$32 agree well with the correlation between
the HBO and lower kHz QPO observed in Z and atoll
sources, hence strengthening our identification.

{\em XTE~J1550$-$564.---\/}A QPO with a frequency that
depends strongly on countrate has been observed from
this newly-discovered X-ray nova. During the initial
phase of the outburst, the QPO frequency increased
from $\simeq 0.082$~Hz to $\simeq 13$~Hz (Cui et al.\
1999; Remillard et al.\ 1999b). During the phase after
the peak of the outburst, the QPO reappeared at a
frequency that drifted from $\simeq 10$ to $\simeq
2.5$~Hz (Remillard et al.\ 1999b). The QPO frequency
was found to vary erraticaly with the spectral
properties and countrate of the source, in a way that
was different between the initial phase and the one
after the peak of the outburst (Remillard et al.\
1999b).

During the initial phase of the outburst, when the
source was turning into the low state, the QPO was at
a frequency $\sim 0.08-0.4$~Hz and was accompanied by
a broader peaked noise component at frequencies $\sim
2-7$~Hz (Cui et al.\ 1999). Figure~2 shows the
correlation between the QPO and noise components at
this phase of the outburst (green open circles at
the lower left corner), which agrees with the
correlation between the HBO and lower kHz QPO
frequencies seen in Z and atoll sources, extrapolated
by more than two orders of magnitude in frequency. 

In both phases of the outburst, when the QPO frequency
was $\gtrsim 5$~Hz, it was sometimes detected
simultaneously with another variable-frequency $\simeq
160-220$~Hz QPO (Remillard et al.\ 1999b). Given the
magnitudes and variability of their frequencies as
well as the identification of the $\sim 0.08-13$~Hz
QPO with the HBO, we tentatively identify the $\sim
160-220$~Hz QPO with the lower kHz QPO seen in
neutron-star sources. Figure~2 shows the correlation
between the frequencies of these two QPOs and compares
it with those of the other neutron-star and black-hole
sources.  The data points that correspond to the
initial and decay phase of the outburst lie on
different branches in Figure~2, in agreement with the
fact that the QPO properties are different between the
two phases of the outburst.

{\em GRS~1915$+$105.---\/}A rich phenomenology of QPO
properties for this source has emerged from recent
{\rm RXTE\/} observations (see, e.g., Morgan et al.\
1997; Trudolyubov, Churazov, \&
Gilfanov\markcite{TCG99} 1999). When the source was in
the very bright state (Morgan et al.\ 1997), a variety
of QPOs with frequencies $\simeq 0.001-10$~Hz as well
as a $\simeq 67$~Hz QPO were observed. These QPOs did
not seem to be related to any of the QPOs observed in
other neutron-star and black-hole sources and may be
related to the activity of GRS~1915$+$105 as a
microquasar. We therefore do not consider these QPOs
in the present study. On the other hand, in its hard
state, the source showed a $\simeq 0.5-10$~Hz QPO with
a frequency that depended strongly on countrate
(Morgan et al.\ 1997; Chen, Swank, \&
Taam\markcite{CST97} 1997; Markwardt, Swank, \&
Taam\markcite{MST99b} 1999b; Belloni et al.\ 1999a).  
When the frequency of this QPO was $\simeq 0.7$~Hz, an
excess of the power-spectral density at $\simeq 7$~Hz
was also evident (Belloni et al.\ 1999a). As before,
we identify the narrow QPO with the HBO and the broad
power-spectral excess with the lower kHz QPO seen in
atoll and Z sources. Figure~2 shows that the
frequencies of these two components are consistent
with the correlation between the HBO and lower kHz QPO
frequencies seen in neutron-star sources.

{\em GRO~1655$-$40.---\/}A $\simeq 300$~Hz QPO has
been detected from this source, always accompanied by
a QPO at $\simeq 10$~Hz (Remillard et
al.\markcite{Retal99a} 1999a). Identifying the $\simeq
300$~Hz QPO with the lower kHz QPO seen in atoll and Z
sources, suggests that the $\simeq 10$~Hz QPO is
similar to the $\simeq 10-50$~Hz QPO seen in
4U~1728$-$34 and in 4U~1608$-$52 (see Fig.\,2). 
Moreover, this identification suggests that the two
QPOs in GRO~1655$-$40 discussed above are similar to
the $\sim 0.08-13$~Hz and $\sim 160-220$~Hz QPOs seen
in XTE~1550$-$564 during the decay phase of its
outburst. However, we stress here that this
identification is very tentative. A number of
additional QPOs and broad peaks are seen in
GRO~1655$-$40, which are not harmonically related to
the ones discussed above (Remillard et al. 1999b). A
careful analysis of the timing properties of this
source is needed, in view of the hypothesis put
forward here, for a more detailed comparison of this
with other sources. 

{\em Other black-hole source.---\/}A number of
additional black-hole sources and black-hole candidate
sources show similar $\simeq 0.1-10$~Hz QPOs: for
example, GS~1124$-$683 (Miyamoto et al.\ 1994; Tanaka
\& Lewin 1995; Takizawa et al.\ 1997; Belloni et al.\
1997; Rutledge et al.\ 1999); LMC~X-1 (Ebisawa,
Mitsuda, \& Inoue 1989); GRO~J1719$-$24 (van der Hooft
et al.\ 1996); 1E~1740.7$-$2942 (Smith et al.\ 1997);
GRS~1758 $-$258 (Smith et al.\ 1997); and
GS~2023$+$338 (Oosterbroek et al.\ 1997). These QPOs
are often accompanied by a power-spectral break at
lower frequencies. In some cases (e.g., GS~1124$-$683
and GS~2023$+$338), broad components at frequencies
higher than the frequency of the QPO are also seen;
however, the properties and frequencies of these
features have not been reported and therefore cannot
be used in the present study.

\subsection{Correlations Between QPO/Noise\\ 
Frequencies}

Figure~2 shows the frequencies of the various types of
variability components detected in Z and atoll sources
as well as in several other neutron-star systems and
in a number of black-hole binaries, as a function of
either the frequency of the lower kHz QPO or the broad
noise component that we tentatively identify as such.
The previously discussed correlations (Psaltis et al.\
1998, 1999) between the frequencies of the upper and
lower kHz QPOs in Z and atoll sources as well as
between the frequencies of the HBO and the kHz QPOs in
Z sources are evident. Moreover, Figure~2 shows that
remarkably tight correlations exist between these
frequencies in the other low-mass X-ray binaries, as
well. 

In particular, there is one correlation extending over
nearly three orders of magnitude in frequency that
appears to depend very little on the properties of the
compact objects and apparently encompasses the HBO and
lower kHz QPO observed in Z and atoll sources, the
$\sim 1-10$~Hz QPO and $\sim 10-100$~Hz noise
component in Cir~X-1, as well as the $\simeq 0.1-1$~Hz
QPO and $\simeq 1-10$~Hz noise component in other
neutron-star and black-hole systems. This relation is
consistent with the data points of 4U~1820$-$30,
4U~1735$-$44, Cir~X-1, 1E~1724$-$3045, GRO~0422$+$32,
and XTE~J1550 $-$564 but statistically only marginally
consistent with the data points of 4U~1728$-$34,
GS~1826$-$36, GRS~1915$+$105, and GX~339$-$4. The
latter does not necessarily imply that the relation is
not consistent with the true frequencies of these
components in the last four sources, as the low
coherence of the broad noise components
($\Delta\nu/\nu$ is of order unity in these sources)
implies that systematic effects such as continuum
subtraction, which we have not taken into account in
our error estimates, dominate the uncertainties in the
measurement of the centroid frequencies.

In addition to the HBO and the upper kHz QPO, the NBO
frequencies of Sco~X-1 are also correlated to the
frequency of the lower kHz QPO (van der Klis 1997).
Further structure is visible in Figure~2 in between
the HBO and NBO correlations, consisting mostly of
QPOs observed in 4U~1728$-$34, 4U~1608$-$52,
XTE~J1550$-$564, and GRO~1655$-$40. These QPOs may
represent the subharmonics of the HBO or demonstrate
the existence of a different type of QPO (see Ford \&
van der Klis 1998 for a discussion). Of course, we
could always hypothesize that there are two types of
lower kHz QPOs. Note here that the very flat
correlation between the HBO and kHz QPO frequencies
seen in Sco~X-1 (see Psaltis et al.\ 1999) may be due
to a transition from one branch to the other, as
Figure~2 suggests. Figure~2 also suggests some other
QPO identifications that have not been previously
made.

\section{DISCUSSION}

We studied the various types of QPOs and broad
noise components observed in neutron-star and
black-hole X-ray binaries. We found that among the
frequencies of these various type of variability
components it is possible to find two that are
tightly correlated in a way that seems to depend
only weakly on the other properties of the sources,
such as the mass, magnetic field, or possibly the
presence of a hard surface in the compact object. 
Such correlations suggest that similar physical
mechanisms may be responsible for some QPOs and
noise components, which can be found over wide
ranges of frequencies and coherences, in Z, atoll,
and possibly even black-hole sources. 

Figure~2 shows these correlations and includes
observations of neutron-star and black-hole sources in
which a QPO (with frequency $\gtrsim 0.1$~Hz) has been
detected simultaneously with at least one more
variability component, either broad-band noise or QPO,
and for which we could obtain reliable frequency
estimates.  There exist other observations of the same
and related sources, where no variability component,
or only one, of the types discussed here is detected.
It is not clear what determines the number and type of
detectable variability components in a given source
and at a given spectral state. However, it appears
that when the variability components we have
identified above are detected in a source, their
frequencies follow one of a small number of
correlations shown in Figure~2. 

In particular, we find indications that the
low-frequency ($\simeq 0.1-100$~Hz) QPOs observed in
some atoll sources, in Cir~X-1 and other neutron-star
sources, as well as possibly in some black-hole
sources may be the same phenomenon as the HBO observed
in Z sources. Moreover, we suggest that the broad
noise components at frequencies $\simeq 1-100$~Hz
observed in Cir~X-1, some neutron-star star sources,
and possibly in some black-hole sources may be the
same phenomenon as the lower kHz QPOs observed in Z
and atoll sources.

We caution here that the identification of broad noise
components in some sources with narrow QPOs in others
relies so far entirely on the tight correlations
between their frequencies shown in Figure~2. No
transition from a narrow QPO peak to a broad noise
component has been observed so far in a single source.
Note, however, that the relative widths of the
variability components identified here as the lower
kHz QPO increase systematically with decreasing
centroid frequency when different sources are
compared: $\Delta\nu/\nu \simeq 0.01-0.5$ in the
$\simeq 200-800$~Hz lower kHz QPOs in Z and atoll
sources, $\Delta\nu/\nu\simeq 0.2-1$ in the $\simeq
20-100$~Hz broad noise component in Cir~X-1, and
$\Delta\nu/\nu\simeq 1-3$ in the $\simeq 1-10$~Hz
broad noise components in the black-hole sources. In
Sco~X-1, the relative width of the lower kHz QPO also
increases systematically with decreasing QPO
frequency, in a way that is consistent with the
decrease of coherence between sources discussed above.
Detection of a transition from a narrow kHz QPO to a
broad, lower frequency noise component would give more
weight to the conjecture put forward here. On the
other hand, detection of one or two narrow kHz QPOs
together with the broad-band noise components at lower
frequencies that follows the tight correlation shown
in Figure~2 will reject our hypothesis.

If confirmed, the detection of HBOs and lower kHz QPOs
over a wide range of frequencies in neutron-star
systems will pose strong constraints on theoretical
models for their nature. A successful model must be
able to explain the presence of the same type of
variability in sources with significantly different
mass accretion rates, with values for the frequencies
and coherences that span over two orders of magnitude.
Figure~2 gives a hint that HBOs and lower kHz QPOs may
have been detected in some black-hole candidates.
Detailed analysis of the properties of these QPOs and
further comparison between the power spectra of
neutron-star and black-hole sources are required to
test such a conjecture.  Confirmation of the
identification of the same type of QPO in both
neutron-star and black-hole sources will challenge
models of such QPOs that depend on the existence of a
hard surface or of an ordered magnetic field around
the compact object. 

\acknowledgements

We are grateful to Eric Ford, Erik Kuulkers, Mariano
M\'endez, Ron Remillard, Rudy Wijnands, and especially
to Robert Shirey for providing us with data in advance
of publication and for many enlightening discussions.
We thank Greg Sobczak, Jeff McClintock, Ed Morgan, and
Ron Remillard for bringing to our attention the QPO
properties of XTE~J1550$-$564. We also thank Deepto
Chakrabarty and Vicky Kalogera for useful discussions
and for carefully reading the manuscript. DP thanks
for their hospitality the members of the Astronomical
Institute ``Anton Pannekoek'' of the University of
Amsterdam, where this work was initiated. This work
was supported in part by a post-doctoral fellowship of
the Smithsonian Institute (DP), by the Netherlands
Foundation for Research in Astronomy (ASTRON) grant
781-76-017 (TB, MvdK), and by several {\it RXTE\/}
observing grants.

\begin{figure*}[ht]
\centerline{
\psfig{file=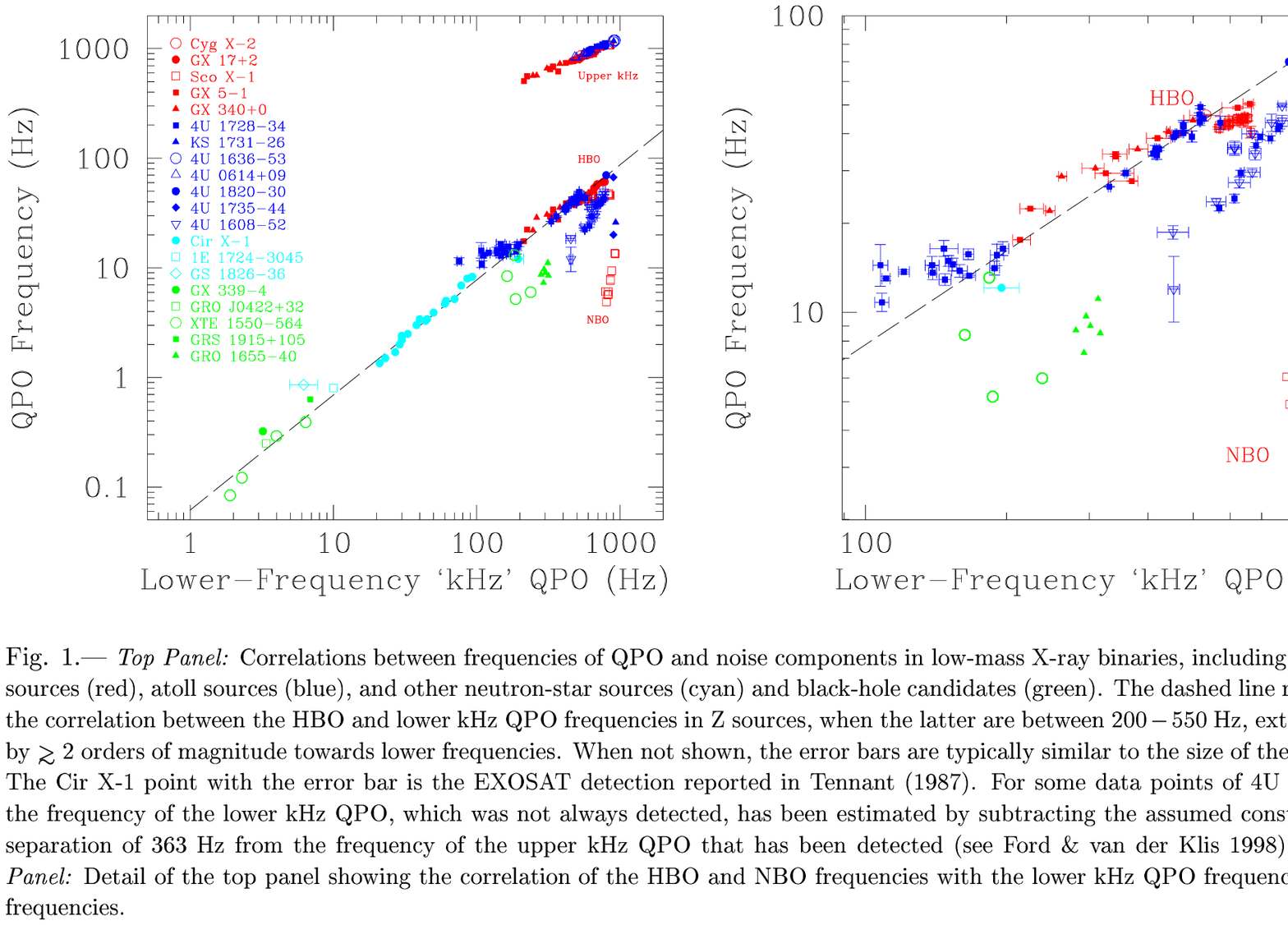,angle=90,height=8.5in,width=6.0in}}
\end{figure*}

\end{document}